\documentclass[conference]{IEEEtran}
\IEEEoverridecommandlockouts
\usepackage{cite}
\usepackage{amsmath,amssymb,amsfonts}
\usepackage{algorithmic}
\usepackage{graphicx}
\usepackage{textcomp}
\usepackage{xcolor}
\usepackage{cite}
\usepackage{siunitx}
\usepackage{graphicx}
\usepackage{tabularx}
\usepackage{float}
\usepackage{subfig}
\usepackage{url}
\usepackage{outlines}
\usepackage{hyperref}
\usepackage{enumitem}
\usepackage{csquotes}
\usepackage{amsmath,amssymb,amsfonts}
\usepackage{algorithmic}
\usepackage{graphicx}
\usepackage{textcomp}
\usepackage{xcolor}

\def\BibTeX{{\rm B\kern-.05em{\sc i\kern-.025em b}\kern-.08em
    T\kern-.1667em\lower.7ex\hbox{E}\kern-.125emX}}
\begin{document}

\title{Systematic Modeling Approach for Environmental Perception Limitations in Automated Driving \\
\thanks{The research leading to these results has received funding from the European Unions Horizon 2020 research and innovation program under the Marie Sk\l odowska-Curie grant agreement No 812.788 (MSCA-ETN SAS). This publication reflects only the authors view, exempting the European Union from any liability. Project website: http://etn-sas.eu/.}
}

\author{\IEEEauthorblockN{Ahmad Adee}
	\IEEEauthorblockA{\textit{Corporate Research,} \\
		\textit{Robert Bosch GmbH,}\\
		Renningen, Germany \\
		Ahmad.Adee@de.bosch.com}
	
	\and
	\IEEEauthorblockN{Roman Gansch}
	\IEEEauthorblockA{\textit{Corporate Research,} \\
		\textit{Robert Bosch GmbH,}\\
		Renningen, Germany \\
		Roman.Gansch@de.bosch.com}
	
	\and
	\IEEEauthorblockN{Peter Liggesmeyer}
	\IEEEauthorblockA{\textit{Software Engineering Institute} \\
		\textit{University of Kaiserslautern,}\\
		Kaiserslautern, Germany \\
		liggesmeyer@cs.uni-kl.de}
}

\maketitle

\begin{abstract}
Highly automated driving (HAD) vehicles are complex systems operating in an open context. Complexity of these systems as well as limitations and insufficiencies in sensing and understanding the open context may result in unsafe and uncertain behavior. The safety critical nature of the HAD vehicles demands to model limitations, insufficiencies and triggering conditions to argue safe behavior.
 
Standardization activities such as ISO/PAS 21448 provide guidelines on the safety of the intended functionality (SOTIF) and focus on the performance limitations and triggering conditions. Although, SOTIF provides a non-exhaustive list of scenario factors that may serve as a starting point to identify and analyze performance limitations and triggering conditions, yet no concrete methodology is provided to model these factors. 

We propose a novel methodology to model triggering conditions and performance limitations in a scene to assess SOTIF. We utilize Bayesian network (BN) in this regard. The experts provide the BN structure and conditional belief tables are learned using the maximum likelihood estimator. We provide performance limitation maps (PLMs) and conditional performance limitation maps (CPLMs), given a scene. As a case study, we provide PLMs and CPLMs of LIDAR in a defined scene using real world data.
\end{abstract}


\section{Introduction}
\label{ch:introduction}
Highly automated driving (HAD) vehicles are complex systems operating in an open context~\cite{burton2020mind}. The complexity and open context nature may result in unsafe and uncertain behavior due to limitations and insufficiencies in sensing and understanding the operational environment~\cite{burton2020mind}. Modeling such limitations and insufficiencies requires the consideration of all possible scenarios and factors influencing the HAD vehicle performance. The international organization for standardization (ISO) published the publicly available specification (PAS), ISO/PAS 21448 road vehicles safety of the intended functionality (SOTIF)~\cite{ISO21448}. The goal of the SOTIF guidelines is to identify the performance limitations and triggering conditions that may lead to potentially hazardous behavior. Specifically, SOTIF is applied to the intended functionality where proper situational awareness is critical to safety and the situational awareness is derived from complex sensors and processing algorithms~\cite{ISO21448}.

Evaluating a perception system (sensor and its processing algorithm) in terms of their limitations, capabilities or inherent uncertainties is not a straightforward task. A perception system cannot be characterized based on a rudimentary set of safety requirements or key performance indicators (KPIs), as the performance of such system depends on many influencing factors. For example, functional performance of a LIDAR based perception system may depend on the spatial distribution of detection, reflection, weather and road conditions.

Modeling the dependencies and influencing factors of the perception system to assess performance limitations and  consequently the relevant uncertainties is important for SOTIF argumentation~\cite{adeeesrel2020}. Such models can provide valuable insights on the functional performance of the system during development. ISO/PAS 21448~\cite{ISO21448} provides a list of such dependencies in terms of scenario factors but does not provide concrete steps to model these scenario factors. 

\begin{figure}
	\centering
	\includegraphics[width=0.65\linewidth]{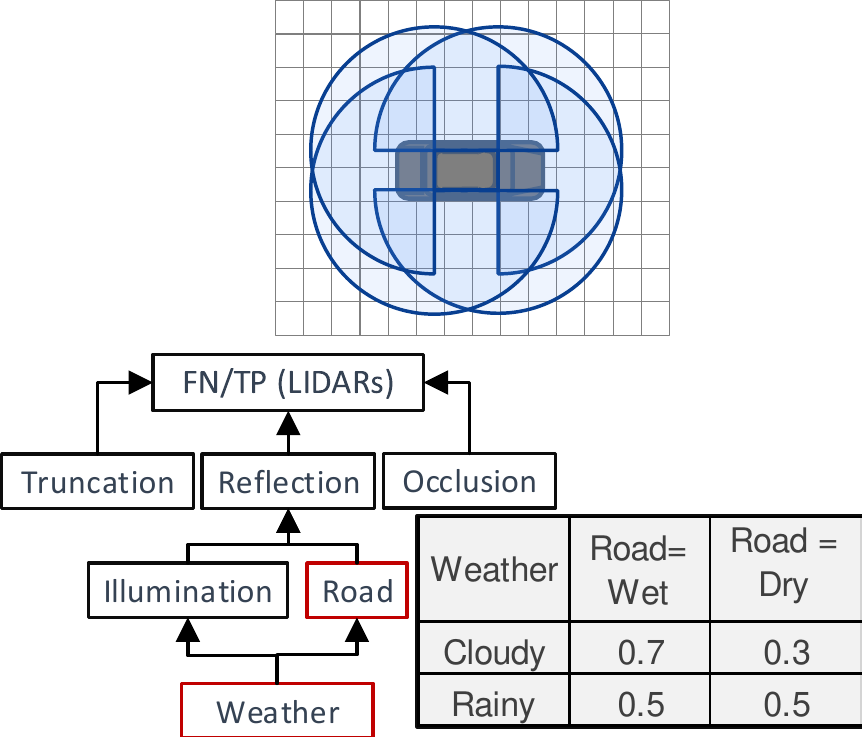}
	\caption{An example of grid map and scene modeling attributed to the cells: LIDAR detections are discretized in grid cell around the field of view. Four LIDARs are attached at the roof of the HAD vehicle for detection. Bottom part shows a Bayesian network along with conditional belief table for $\Pr(Road\mid Weather)$. } \label{fig:grid_ex}
\end{figure}

Probabilistic graphical models (PGMs)~\cite{koller2009probabilistic} in general and Bayesian networks (BNs)~\cite{pearl2014probabilistic} in particular have rapidly gained popularity in the dependability research~\cite{weber2012overview},~\cite{cai2018application}. The BN is a directed acyclic graph (DAG) that consists of nodes and edges. Every node is a random variable $(X_1,\dots,X_n)$, which represents an element of the system or its context. The edges represent a directed relationship between two nodes and run from the parent node $(pa)$ towards the child node $(ch)$. Together, nodes and edges represent the structure of the probabilistic network (Fig.~\ref{fig:grid_ex}). The strength of these dependencies are governed by conditional probability distributions $\Pr(ch\mid pa)$~\cite{koller2009probabilistic}. Mathematically, the BN can be written as follows.
\begin{equation}
	\Pr (X_1,\dots,X_n) = \prod_{i}^{n} \Pr(X_i\mid pa(X_i))
\end{equation}
BN is effective in modeling uncertainty and probability reasoning of a system. It exploits the dependence relationship through the local conditions in the model to perform uncertainty analysis for prediction, classification and causal inference of influencing factors.

In this publication, we formulate a model using BN for known triggering conditions and performance limitations in a given scene. A human expert provides the SOTIF relevant scenario factors and models the causal relations among them using a BN structure. We perform parameter learning of BN to quantify the dependencies in the model. In order to explain the performance limitations and triggering conditions effects on SOTIF, posterior probability analysis and causal inference is conducted. We construct performance limitation maps (PLMs) and conditional performance limitation maps (CPLMs) using these analyses. Causal inference identifies the most contributing influencing factors on performance. Together, PLMs, CPLMs and causal inference provide valuable insights on the SOTIF. This may help the analyst in the safety case generation, identification of the performance limitations, generation of targeted test, validation and verification campaigns and influencing factors, which in turn can help in defining refinement measures. Summarizing, we provide the following contributions.
\begin{itemize}
	\item We introduce a method to model known triggering conditions and performance limitations in a scene.
	\item We introduce PLMs as the representation of SOTIF metric.
	\item We introduce CPLMs to quantify the effects of triggering conditions and influencing factors on SOTIF.
	\item We implement the methodology and provide PLMs and CPLMs of LIDARs case study while utilizing real world data.
\end{itemize}

The publication is structured as follows: Sec.~\ref{ch:methodology} presents the proposed methodology. Sec.~\ref{ch:setup} briefly describes the setup used for data acquisition. Sec.~\ref{ch:Implementation} provides the application of proposed methodology on LIDAR perception. In sec.~\ref{ch:Results}, results of the implementation are evaluated. Sec.~\ref{ch:evaluation} provides the evaluation of the approach and robustness of the results. Sec.~\ref{ch:RelatedWork} provides an overview on the state of the art. Finally, in sec.~\ref{ch:conclusion} we discuss conclusion and future work.
\section{Proposed Methodology}
\label{ch:methodology}
\begin{figure}
	\centering
	\includegraphics[width=0.7\linewidth]{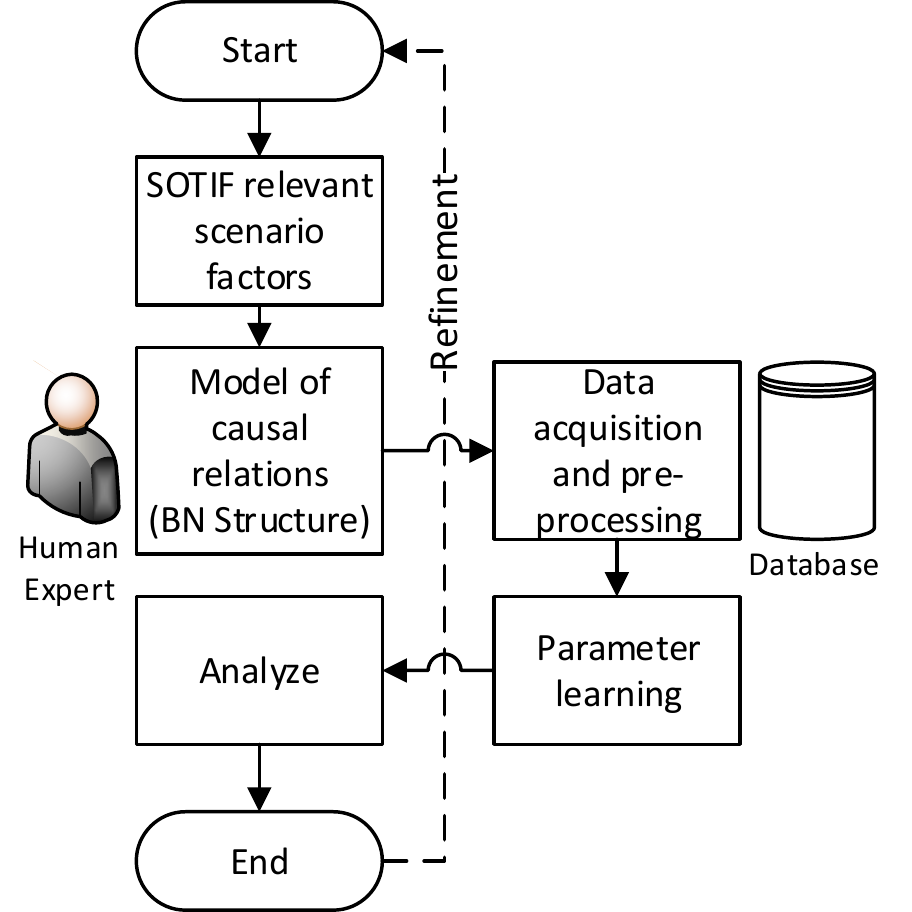}
	\caption{Flowchart describing the flow of the proposed methodology. SOTIF relevant scenario factors and expert knowledge are encoded into scene model defined by the BN structure. Data is gathered accordingly and learning of parameters is performed.} \label{fig:flowchart}
\end{figure}

We introduce a modeling methodology using BN to identify, model and quantify performance limitations as well as triggering conditions in a scene. The experts provide the structure of BN while the conditional belief tables (CBTs) are learned from real sensor data. Fig.~\ref{fig:flowchart} shows the flowchart of the methodology we adopt in this publication. A detailed explanation of the steps proposed in the flowchart (Fig.~\ref{fig:flowchart}) follows. 
\subsection{SOTIF Relevant Scenario Factors}\label{subsec:SRSF}

The first step towards modeling relevant SOTIF scenario factors is the identification of performance limitations and triggering conditions in a given scene~\cite{ISO21448}. 
SOTIF provides a dynamic element and scenery centric non-exhaustive list of scenario factors~\cite{ISO21448}. Although this list can be a starting point, yet identification of triggering conditions and performance limitations is dependent on many other aspects including the context of driving, perception system in question and existing setup among other. For example, consider the following two descriptions.
\begin{enumerate}
	\item Context: Highway, Perception: Radar based, Studied behavior: False Positives. 
	\item Context: Urban, Perception: LIDAR based,  Studied behavior: Position Trueness.
\end{enumerate}

Both description may lead to different scenario factors. In the former, the human expert might be interested in steel bridges, tin cans and other such instances while in the latter the factors of interest may include weather conditions, exhaust gases and reflections.
The process is similar to hazard identification and risk assessment (HARA) from ISO 26262~\cite{iso26262}, but does not explicitly considers malfunctioning behavior of components. It assesses the intended functionality of HAD vehicle functions especially where situational awareness is critical to safety.
We utilize the scenario factors from ISO/PAS 21448~\cite{ISO21448} as well as expert opinion, previous data and existing setup (constraint on data acquisition and/or data labels) to model the scene in our methodology (Fig.~\ref{fig:flowchart}). 

SOTIF related undesired behavior (e.g. braking when not required and vice versa) may originate from FP and FN detections~\cite{ISO21448}. Since we are more focused at the perception level of the functionality, we only consider FN and true positive (TP) of the perception system.
The overall methodology we define in this publication, is however generic and can be applied to complete functional chain (sense, plan, decide and act) of the system under study. The choice of undesired behavior is highly dependent on the system under study, the scene model and metrics that can support the safety case.
Apart from TP and FN, SOTIF related undesired behavior such as FP, positional error, contour matching, classification as well as regression quality can also be modeled to assess the performance limitation and the effects of triggering conditions on the functional performance.
As an example, for the second case in which LIDAR based perception system is analyzed in the context of urban driving, the expert may provide the following factors.
\begin{itemize}
	\item \textbf{Occlusion}: In urban driving, there may be a relatively higher probability of occlusion occurrence as parked cars, trees may occlude objects.
	\item \textbf{FN/FP} rate: The overall FN/FP rate in the urban context of driving.
	\item \textbf{Weather} conditions: Different weather may effect the LIDAR performance.
	\item \textbf{Reflection} from objects: Reflection from different objects (buses windows) effects the FP rate.
	\item \textbf{Illumination}: Higher illumination may increase the reflection from objects.
\end{itemize}
The above-mentioned factors are non-exhaustive. Scenario factors are provided and refined based on the expert opinion and ability for data acquisition. The resulting factors then can be used to model the causal relation.
\subsection{Model of the Causal Relation}\label{subsec:methodcausalrelation}
Modeling the qualitative and casual relations amongst the scenario factors, triggering conditions and performance limitations is a significant component of this methodology. We utilize BN structure for this purpose. Traditionally, the BN structure modeling is based either on the expert knowledge~\cite{flores2011incorporating} or on the learning from data (structure learning)~\cite{scanagatta2019survey}. 
However, in structure learning from data the number of graph candidates grow exponentially with the number of variables in the data~\cite{kabli2007chain}. Discerning true graph by using observational data alone from other graphs that model the same set of conditional independencies is also challenging. Due to these challenges, we opt for the former technique in this work.

Scene description, which include SOTIF relevant scenario factors and corresponding undesired behavior(s) constitute the nodes of the BN structure. As a first step towards derivation of the structure, the experts establish hierarchical dependencies between undesired behavior, triggering conditions of the scene, and performance limitations and provide propositions e.g. the proposition $p1$ : high occlusion may result in higher FNs. We then construct BN with arcs representing the dependencies and nodes representing the undesired behavior, triggering conditions and performance limitations derived from these propositions e.g., the proposition $p1$ is modeled as an explicit node (Fig. ~\ref{fig:BN_Structure}). The resulting BN structure asserts that a child node is governed by a causal mechanism that probabilistically determines its value based on the values mechanism of its parents~\cite{koller2009probabilistic}. The stochastic attribute of such models helps modeling aleatory uncertainty~\cite{gansch2020}.
\subsection{Data Acquisition and Pre-processing}\label{subsec:methoddataacc}
Dataset $\mathcal{D}$ acquired and utilized in our methodology consists of fully observed instances of the network variables.
\begin{equation}\label{eq:dataset}
\mathcal{D} = \xi[1] \dots \xi[M]
\end{equation}
Where $\xi[.]$ represents a data instance and $M$ represents the number of instances in $\mathcal{D}$. 

We calculate SOTIF related undesired behavior for each data instance, if the undesired behavior is not labeled. For example, data instances may not be labeled with FNs. 
However, this is an ad-hoc step for data processing that may or may not be required, depending upon the available dataset.

In order to fully grasp the effects of SOTIF relevant scenario factors (conditional dependencies in BN) and performance limitations around the HAD vehicle, we discretize the spatial distributions of detections in a grid map (Fig.~\ref{fig:grid_ex}). Modeling spatial distribution of triggering conditions and performance limitations in a grid map is important for the following reasons.
\begin{enumerate}
	\item Scenario factors are spatially distributed e.g. in a weather situation involving dense fog the FN rate of the grid cells farther from the HAD vehicles will be different than the nearer ones, for some perception systems.
	\item Safety criticality is variable around the vehicle in the sense that events nearer to the HAD vehicle are generally considered more critical.
\end{enumerate}
Data instances thus can be spatially associated around the HAD vehicle to fully associate the observed instances with their respective detection points in space. In this way, a grid map is created around the HAD vehicle to represent SOTIF relevant perception metrics/properties (Fig.~\ref{fig:grid_ex}). For the construction of grid map, a coordinate system (e.g. Cartesian or polar) is selected as well as the grid size. Each grid cell is then represented by a separate BN and its corresponding CBTs (Fig.~\ref{fig:grid_ex}). The structure of each BN is kept constant in this work.

Suppose the data instances are distributed into $\mathcal{N}$ number of grid cells (thus $\mathcal{N}$ number of BNs) based on the Cartesian $(x,y)$ or polar $(r,\theta)$ coordinates of detection. The dataset (Eq.~\ref{eq:dataset}) can be re-written as. 
\begin{equation}
\mathcal{D}^k = \xi^k[1] \dots \xi^k[M^k] \forall k \in \mathcal{K}
\end{equation}
Where $\mathcal{K}$ is a set as follows.
\begin{equation}
	\mathcal{K} = \{ 1,2,\dots,\mathcal{N}\}
\end{equation}
Here $k$ represent $k^{th}$ grid cell and BN.
\subsection{Parameter Learning}
Once BN structure (Sec.~\ref{subsec:methodcausalrelation}) is determined and corresponding data is acquired (Sec.~\ref{subsec:methoddataacc}), the CBTs can be learned. We determine the CBTs and thus the strength of the dependencies by utilizing the maximum likelihood estimator (MLE)~\cite{koller2009probabilistic}. We perform non-parametric learning, not assuming prior probabilities. Given a variable $X$ with parents $\textbf{U}$, we will have a parameter $\theta_{x\mid \textbf{u}}^k$ for each combination of $x^k \in Val(X)$ and $\textbf{u}^k \in Val(\textbf{U})$ for a CBT. The likelihood function for such case is as follows.
\begin{equation}\label{eq:likelihood}
	L_X (\theta_{X \mid U}^k: \mathcal{D}^k) = \prod_{m} \theta_{x[m]\textbf{u}[m]}^k = \prod_{\textbf{u} \in Val(\textbf{U})} \prod_{x \in Val(X)} {\theta^k}_{x \mid \textbf{u}}^{M^k[\textbf{u},x]}
\end{equation}
Here $\theta^k_{x \mid \textbf{u}}$ represents the parameter to be learned, $k$ represents the $kth$ BN around the HAD vehicle and $m$ represents the $m^{th}$ data instance in the dataset. 
Maximizing the likelihood function from Eq.~\ref{eq:likelihood} results in the learned parameter.
\begin{equation}\label{eq:maxlikelihood}
\theta_{x \mid \textbf{u}}^k = \frac{{M^k[\textbf{u},x]}}{{M^k[\textbf{u}]}}
\end{equation}
Here ${M^k[\textbf{u},x]}$ represents the combined occurrence of $u$ and $x$ for the $k^{th}$ BN. Eq.~\ref{eq:maxlikelihood} defines the MLE.
\subsection{Refinement}
The aim of refinement steps is to improve the BN (both structure and CBTs), so that exhaustive and complete models for SOTIF can be produced. We believe that this a hybrid approach (involving experts while partially automating the approach) may provide the most suitable results. Every step explained in the previous sections and depicted in the Fig.~\ref{fig:flowchart} is subject to iterative refinement based on the analyses and obtained results. This includes additions/deletion of scenario factors, restructuring of the BN structure or acquisition of more data.

\section{Experimental Setup}
\label{ch:setup}
The experimental setup consists of two Hesai Pandar 64 and two Velodyne Ultra Puck VLP-32C LIDAR sensors installed on the roof corners of a car (Fig.~\ref{fig:grid_ex}). The recorded data consists of different labels including bounding boxes, pose, visibility state and vehicle activity among others surrounding $\ang{360}$ of the HAD vehicle. The data was collected mostly on the highway in the nearby regions of Stuttgart, Germany. The data consists of around twenty thousand instances. 
A deep neural network (DNN) was trained and used as the processing algorithm. 
Two experts  from the field with substantial experience in the LIDAR based perception systems provided their opinions on LIDAR insufficiencies, triggering conditions and limitations based on the observations in the data acquisition process and experiences with the LIDAR based perception systems. 

\section{Implementation}
\label{ch:Implementation}
In this section, we demonstrate the application of our methodology on the LIDAR sensing dataset discussed in the previous section.
\subsection{SOTIF Relevant Scenario Factors}
The experts provide	 different factors that may effect the LIDAR perception system performance (Sec.~\ref{subsec:SRSF}). Based on the expert inputs, SOTIF scenario factors and availability of data acquisition setup, we include the nodes shown in Fig.~\ref{fig:BN_Structure} as SOTIF relevant scenario factors.

Truncated or occluded objects may only produce sparse point measurements. The occlusion and truncation both representing the visibility state of an object is defined analogously to the KITTI benchmark~\cite{Fritsch2013ITSC}. 
Weather conditions may effect the road conditions and light intensity that in turn can effect reflection on road. Especially heavy rain may cause flooding on road, which in turn can decrease the TP in detections~\cite{goodin2019predicting}.

We use FN and TP rate to represent the SOTIF measure as they are considered adequate measures for SOTIF analysis~\cite{ISO21448}.
\subsection{Model of the Causal Relation}
\begin{figure}[h]
	\centering
	\includegraphics[width=0.9\linewidth]{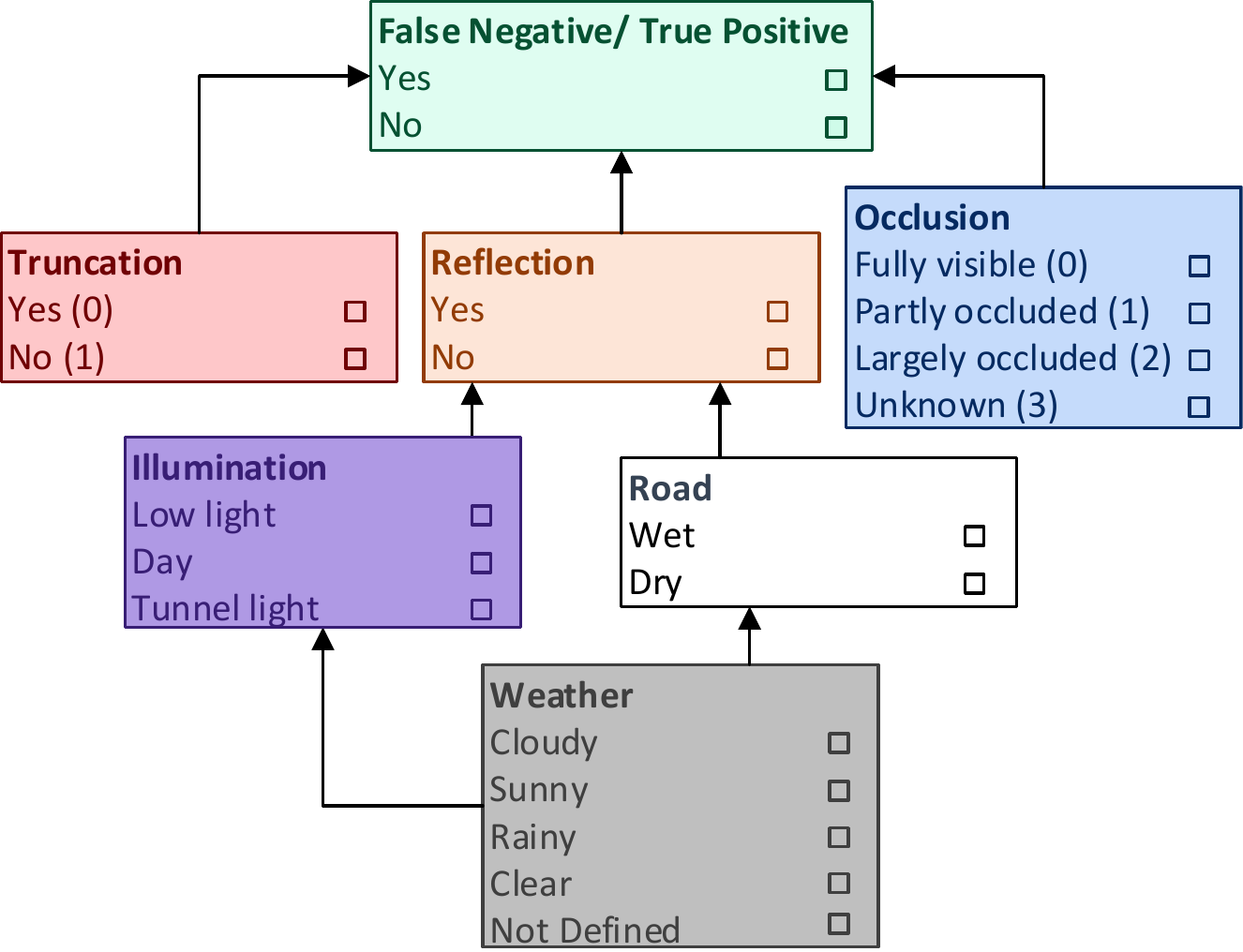}
	\caption{BN based on the SOTIF relevant scenario factors and expert knowledge describing the causal structure used in our implementation. False negative and true positive are selected alternatively. Color coding is provided to support the subsequent figures of the analysis.} \label{fig:BN_Structure}
\end{figure}

Based on the propositions from the previous section, the BN structure is developed. The effects discussed in the previous section can be encoded in the following simple propositions.
\paragraph{Proposition 1} \textbf{Truncation} and \textbf{occlusion} in detection may influence FN and TP.
\paragraph{Proposition 2} \textbf{Weather} conditions may effect road conditions and scene illumination, which in turn can effect the TP/FN rate.
\paragraph{Proposition 3} \textbf{Road condition} and scene \textbf{illumination} can effect \textbf{reflection} in the scene, which in turn can effect the TP/FN rate.\\
The resulting BN structure is shown in Fig.~\ref{fig:BN_Structure}. The BN model contains seven nodes. Once the CBTs are established, the BN can be updated with new information. 
\subsection{Data Acquisition and Pre-processing}
The dataset we use in this paper provides detection and corresponding ground truth (separate datasets). One Bounding box is labeled for each object detection and its corresponding ground truth. All relevant nodes (Fig.~\ref{fig:BN_Structure}) are labeled except TP and FN. In order to evaluate TP and FN for each data instance we use mean squared error (MSE).
\begin{equation}\label{eq:MSE}
MSE = \frac{1}{n}\sum_{i=0}^{n}(Y_i - \hat{Y_i})^2
\end{equation}
Where $n$ represents number of samples, $Y_i$ represents the ground truth and $\hat{Y_i}$ represents the detection. We execute Eq.~\ref{eq:MSE} for individual detections and find a corresponding sample in the ground truth using $x$ and $y$ values. All those data instances from detection which return a data instance from ground truth are considered TP, while all those data points from ground truth that do not return a corresponding value from detection are considered to be FN. Data instances with $|x|>140$ meters and $|y|>50$ meters are not considered as the defined optimal range of LIDAR.

Resolution of grid cells in the grid-map is an interesting aspect as it directly influences the TP and FN rate. This phenomenon is analogous to discretization of continuous spatial distribution as a BN node~\cite{vinh2012data}. As coarsening may result is less precise and accurate while refinement may result in precise and less accurate CBTs~\cite{fenton2018risk}, a well thought discretization is required. Both, static and dynamic discretization can be performed in this regard~\cite{fenton2018risk}. Based on the availability of data for each cell and complete representation of all the nodes of the BN structure (Fig.~\ref{fig:BN_Structure}), we use $x=20$ and $y=10$ meters accordingly, in this publication.
\subsection{Parameter Learning}
We perform parameter learning for individual BN (representing a grid cell) using its corresponding data instances and Eq.~\ref{eq:maxlikelihood}. 
After the establishment of the BN structure and learning of the distribution parameters (CBTs), the BN can be used as an effective tool for analysis and estimation. The resultant PLMs and CPLMs can be used as metrics to assess the SOTIF, identify triggering conditions as well as provide safety cases and validation targets.
\subsection{Refinement}
We provide preliminary refinement steps in the results (Sec.\ref{ch:Results}).
\section{Results}
\label{ch:Results}
In this section, we present the results obtained by applying our methodology. 
\subsection{Performance Limitation Map}
\begin{figure}
	\centering
	\includegraphics[width=\linewidth]{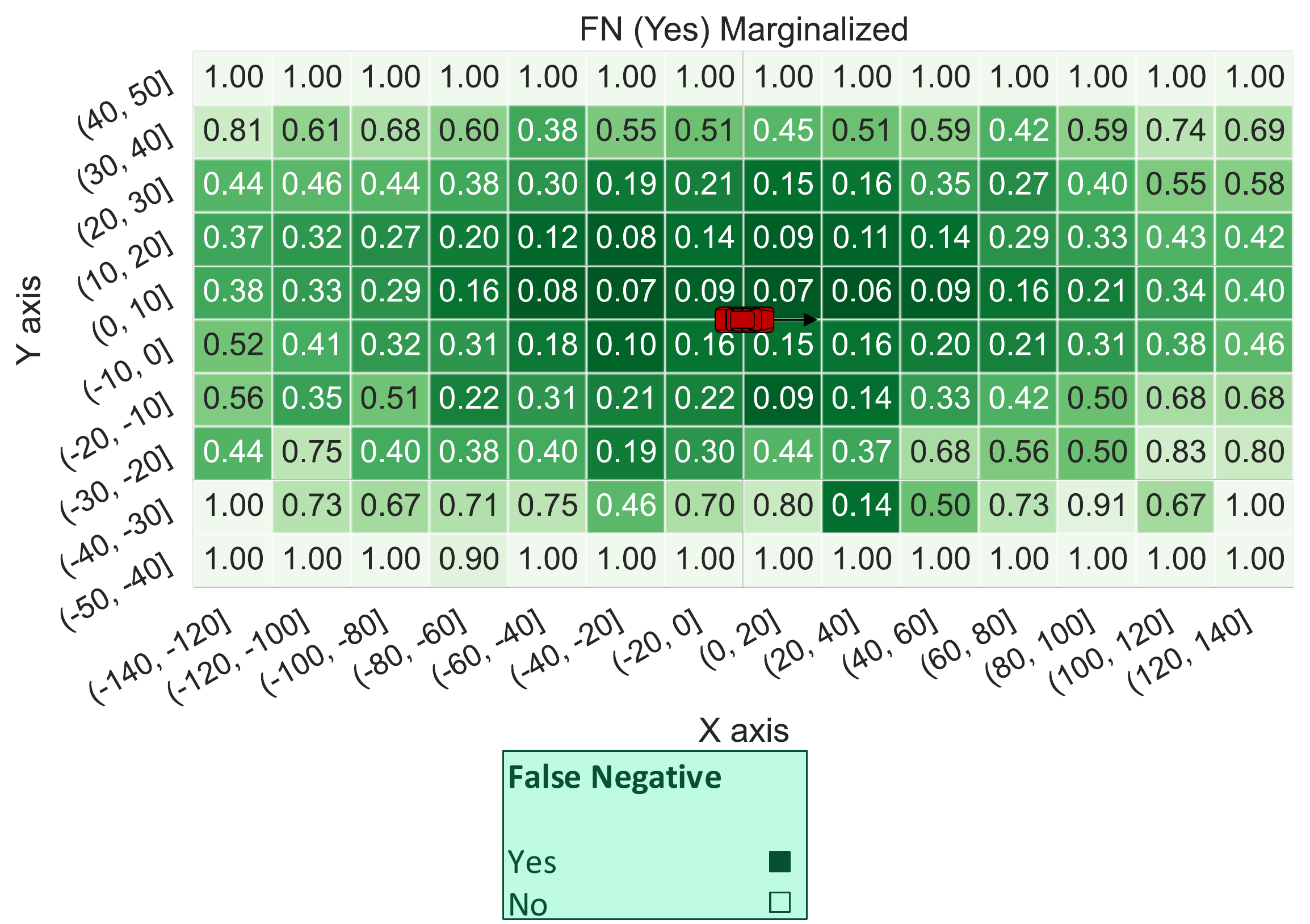}
	\caption{Performance limitation map for FN rate in the described scene and available data used for learning. Better performance of LIDAR is be observed near the HAD vehicle.} \label{fig:FN_Yes_Marginalized}
\end{figure}

\begin{figure}
	\includegraphics[width=\linewidth]{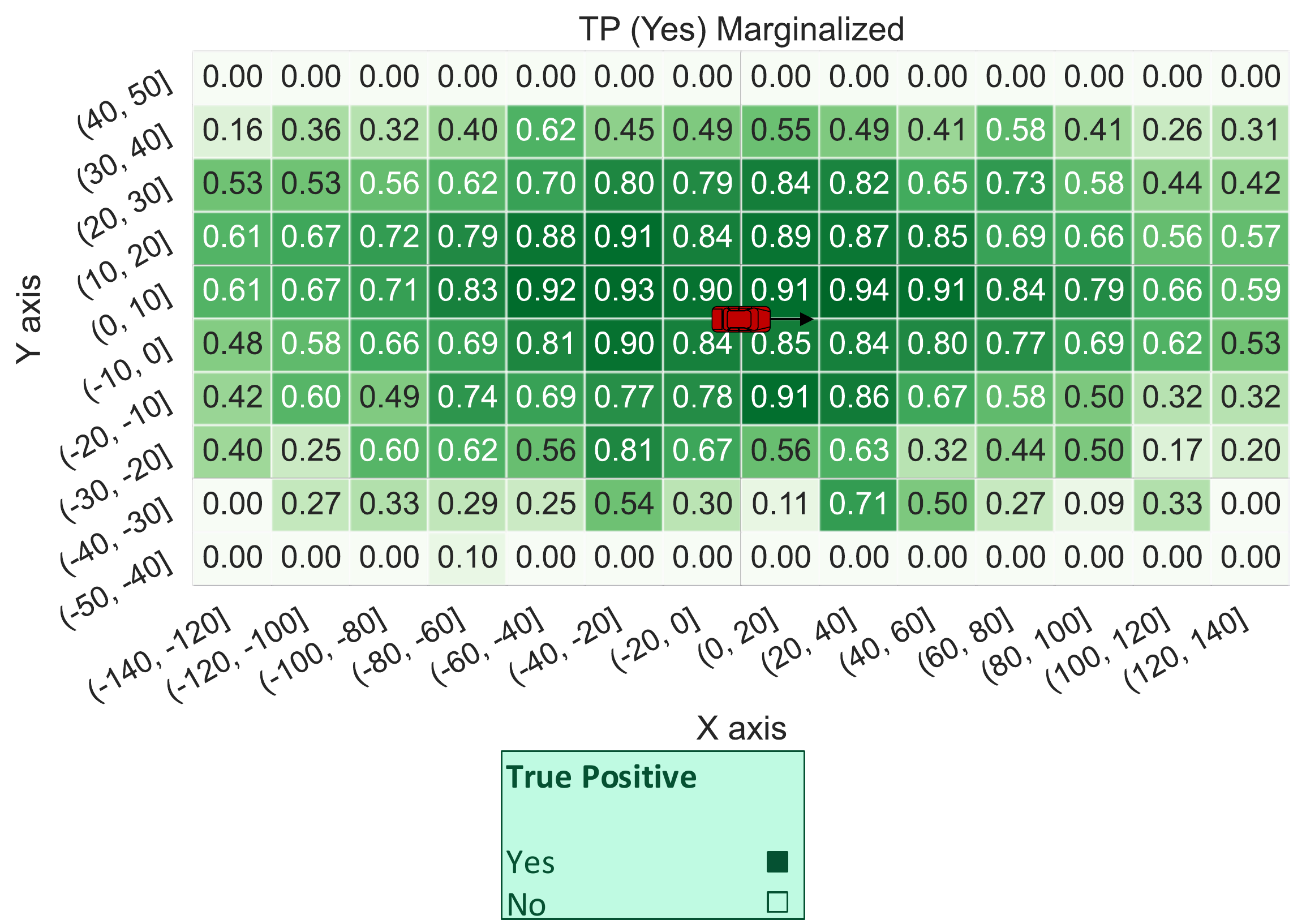}
	\caption{Performance limitation map for TP in the described scene. Better performance of LIDAR is observed near the HAD vehicle.} \label{fig:TP_Yes_Marginalized}
\end{figure}

Grid maps pertaining to TP and FN are shown in Fig.~\ref{fig:FN_Yes_Marginalized},~\ref{fig:TP_Yes_Marginalized}. Essentially PLMs represent the marginalized posterior probability distributions $(\Pr(ch))$ of specific nodes. The heat map reference is reversed for TP to keep the same color for undesired probabilities. 
We can infer the following conclusions.
\begin{enumerate}
	\item We observe better detection capabilities near the HAD vehicle.
	\item Both TP and FN rate are symmetrically distributed across $X$ and $Y$ $axes$ with slightly higher FN rate (and lower TP rate) in front and on the right side of the HAD vehicle.
\end{enumerate}
By using the PLMs (Fig.~\ref{fig:FN_Yes_Marginalized},~\ref{fig:TP_Yes_Marginalized}), the uncertainty of the scene can be represented with a PLM, which can be expressed as a quantitative evaluation of the safety of intended functionality (SOTIF) for a given scene and system under consideration.
%
\subsection{Conditional Performance Limitation Map}
\begin{figure}
	\centering
	\includegraphics[width=\linewidth]{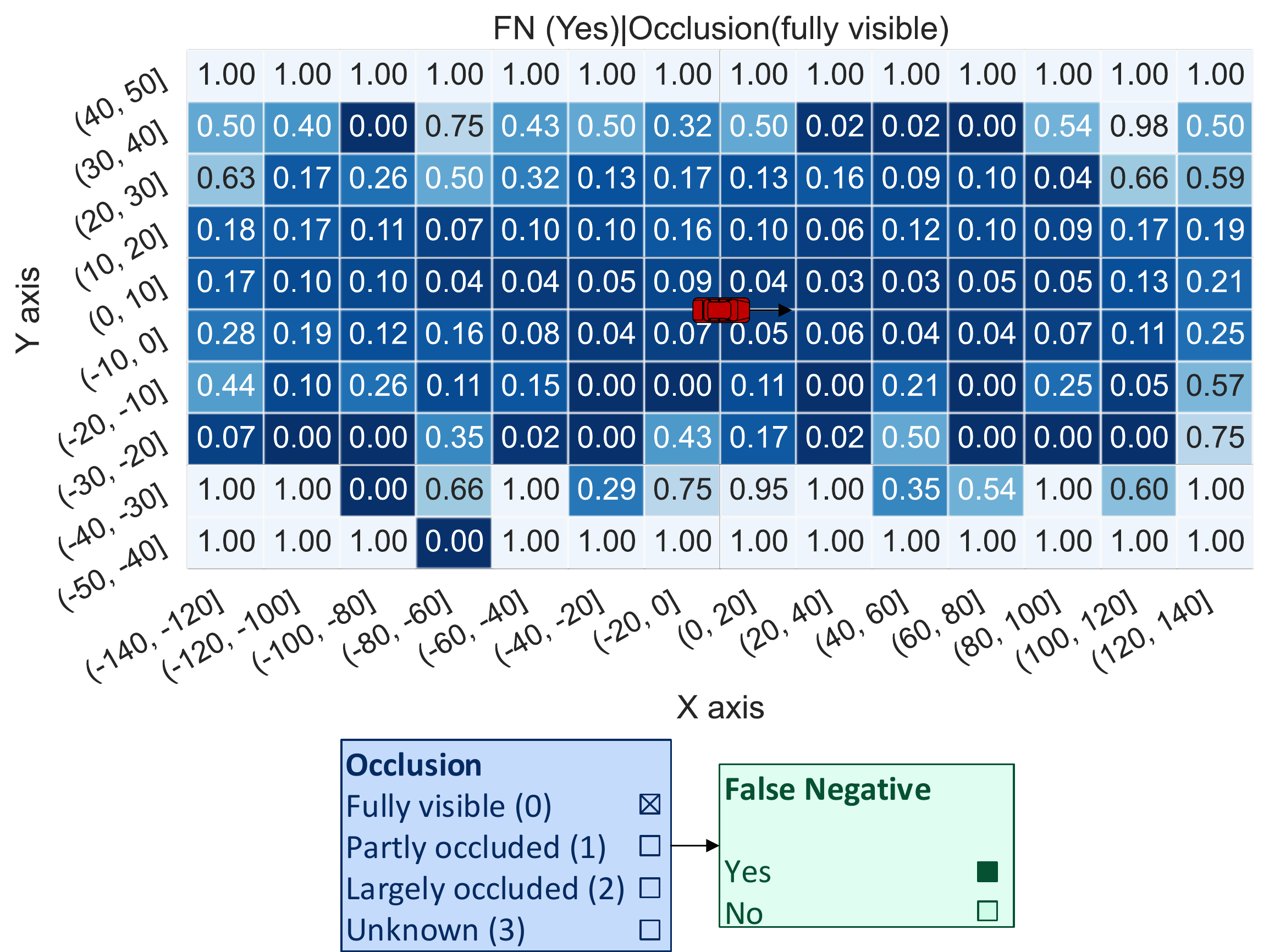}
	\caption{Conditional performance limitation map (CPLM) for FN (Yes) conditioned on Occlusion (fully visible) in the described scene. CPLM for FN describes that occlusion (fully visible) scenes may not causes higher FN rate.} \label{fig:FN_Yes_Occlusion_0}
\end{figure}

\begin{figure}
	\centering
	\includegraphics[width=\linewidth]{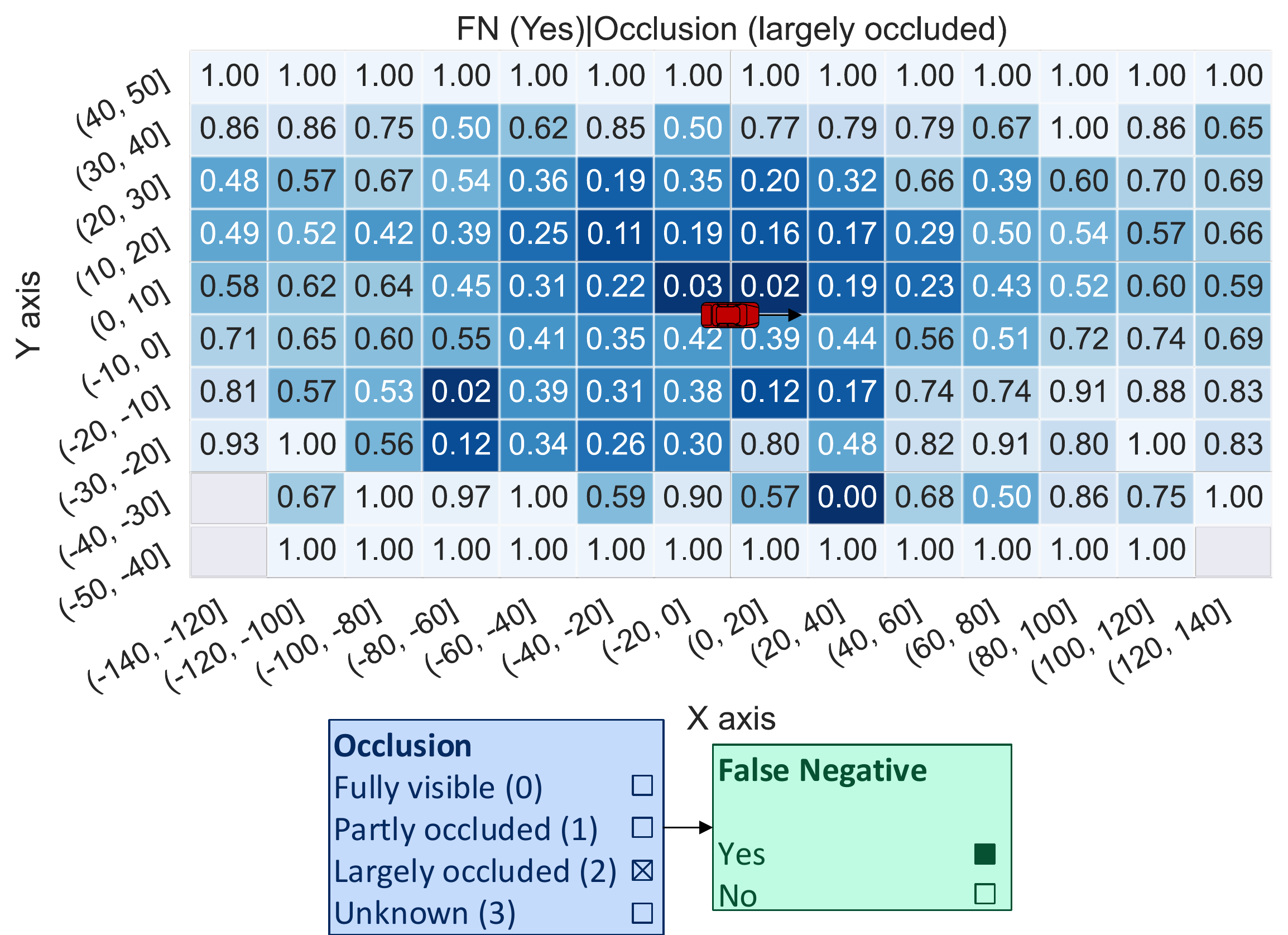}
	\caption{Conditional performance limitation map (CPLM) for FN conditioned on Occlusion (largely occluded) in the described scene.  CPLM for FN describes a higher FN rate for occlusion (largely occluded) scenes.} \label{fig:FN_Yes_Occlusion_2}
\end{figure}

Another interesting analysis result comes in the form of a CPLM, conditioned on individual or multiple nodes of the scene. This corresponds to conditional probability of a child $(ch)$ node given its parent(s) $(pa)$ node $(\Pr(ch\mid pa))$. The $pa$ can be selected individually or in combination. In the light of ISO/PAS 21448, it can be seen as how triggering conditions influence the performance~\cite{ISO21448}. We provide CPLMs of FN conditioned on occlusion (Fig.~\ref{fig:FN_Yes_Occlusion_0},\ref{fig:FN_Yes_Occlusion_2}). Evidently, $occlusion$$=$$largely$ $occluded$ scenes have higher probabilities of FNs than $occlusion$$=$$fully$ $visible$ scenes. We can infer the following conclusions.
\begin{enumerate}
	\item Largely occluded scenes have higher FN rate than fully visible scenes for LIDARs, given the data.
	\item The average $\Pr(FN\mid Occlusion)$ rate is symmetrically distributed across $X$ and $Y$ axes with slightly higher FN rate in front and on the right side of the HAD vehicle.
\end{enumerate}
\subsection{Causal Inference}
\begin{figure}
	\centering
	\includegraphics[width=\linewidth]{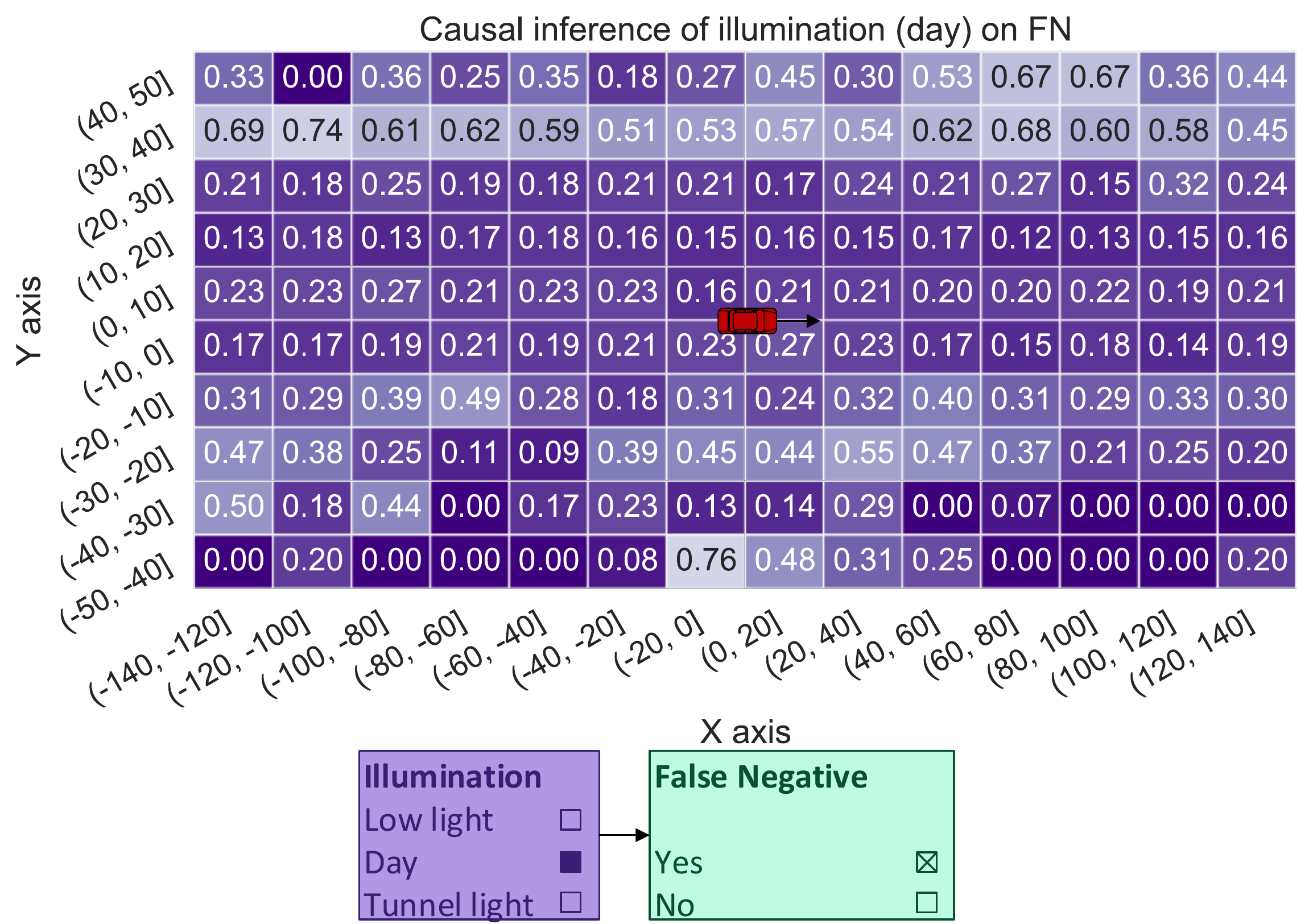}
	\caption{Causal inference map for FN when illumination (day). Illumination state "day" has varied effects on the detection.} \label{fig:Causal_influence_of_illumination_day_on_FN}
\end{figure}

\begin{figure}
	\centering
	\includegraphics[width=\linewidth]{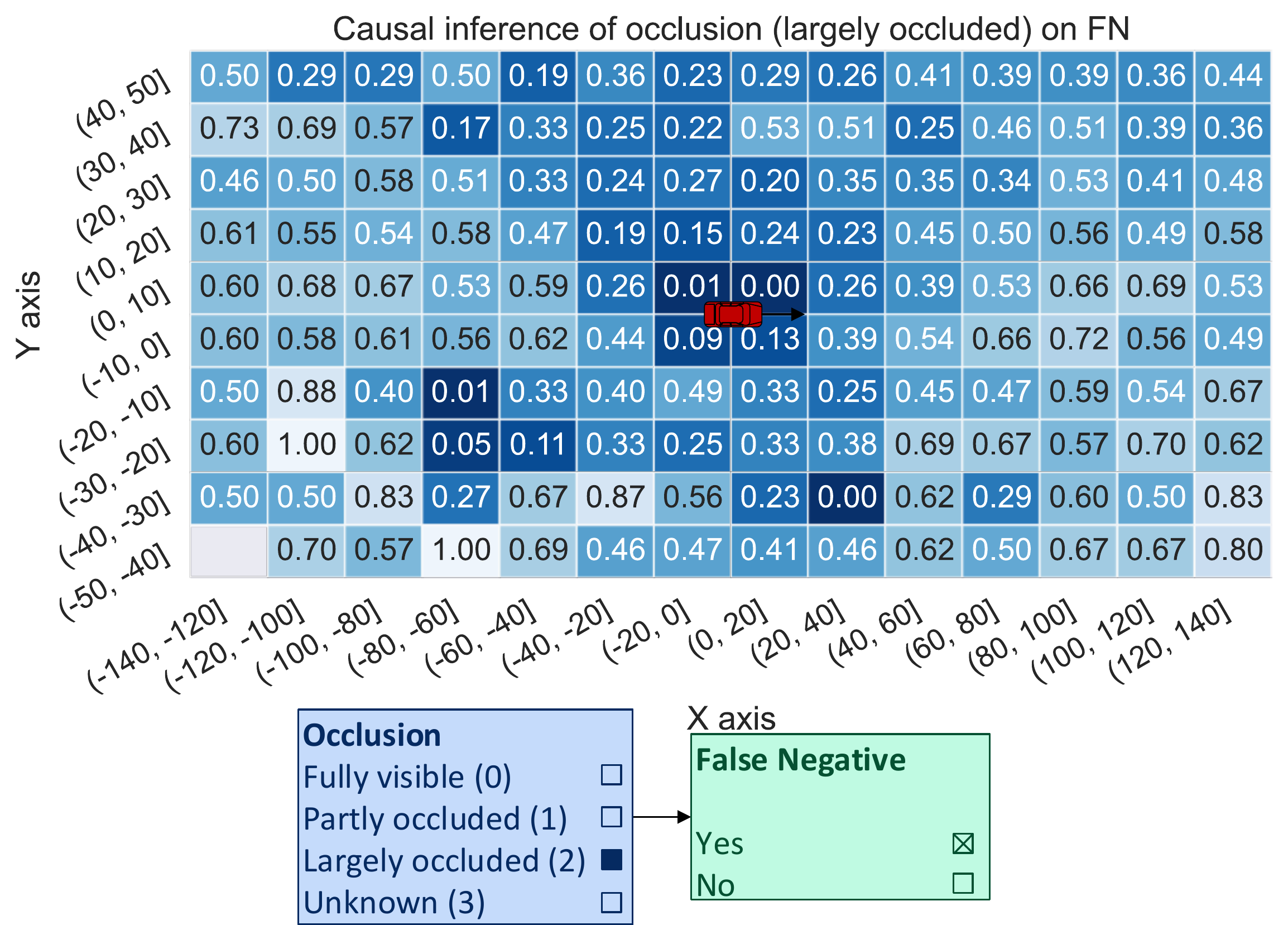}
	\caption{Causal inference map for FN when occlusion (largely occluded). Largely occluded states have considerable effects on the detections away from the HAD vehicle.} \label{fig:Causal_influence_of_occlusion_2_on_FN}
\end{figure}

The strength of BN to provide backward propagation of evidence~\cite{koller2009probabilistic} provides an added advantage of performing causal inference. In other words, casual inference estimates the strength  of a parent node on the child node or any other node in the structure$(\Pr(pa\mid ch$ $or$ $any))$. For example, given the BN, consider the following query.
\begin{paragraph}
	{Query} What causes the FN rate?
\end{paragraph}\\
This query can be answered by setting the FN in the grid map to $Yes$. 
Fig.~\ref{fig:Causal_influence_of_illumination_day_on_FN},\ref{fig:Causal_influence_of_occlusion_2_on_FN} shows the causal inference of $\Pr(Illumination$$=$ $Day \mid FN=Yes)$ and $\Pr(Occlusion= largely$ $occluded \mid FN=Yes)$ respectively. We can infer the following conclusions. 
\begin{enumerate}
	\item  $Occlusion=largely$ $occluded$ has higher impact than $illumination=day$ on FN.
\end{enumerate}
Such results may directly indicate the relevant triggering conditions of performance limitations and may provide a way forward for informed improvement in the design of the system from the SOTIF viewpoint. For example, based on the CPLM (Fig.~\ref{fig:FN_Yes_Occlusion_2}) and a benchmark for FN rate for each cell (given a constant severity and controllability), we may infer that $largely$ $occluded$ scenes are risk factors to SOTIF. However, defining a benchmark is out of the scope of this work.
\subsection{Refinement}
Some of the refinement steps proposed by the experts are.
\begin{itemize}
	\item More data is required for truncation node in order to establish or negate a causal relation.
	\item Abrupt zero values in regions where surrounding grid cells have relatively higher values (Fig.~\ref{fig:FN_Yes_Occlusion_0}) are observed for occlusion ($fully$ $visible$). These cells require further analysis and data instances for robust results. 
\end{itemize}
The refinement steps are non-exhaustive and provision of an exhaustive list of steps is out of the scope of this work.
\section{Evaluation}\label{ch:evaluation}
We perform evaluation of our learned PLM using the test dataset. We predict the FN by essentially setting the weather, occlusion, road and reflection states from the test dataset as evidence and predicting the FN. The results are then compared with the FNs computed by using Eq.~\ref{eq:MSE}. Tab.~\ref{tab:evaluation} shows the results of our evaluation. We observe that substantial accuracy in the results can be achieved. However, like any other data oriented implementation, measuring the true underlying parameter distribution (or CBTs) is a challenging task. In general, a parameter learning algorithm for BN extracts the joint relative frequency if they have conditional relation in their structure e.g. $X\mid Y$~\cite{koller2009probabilistic}. As the real CBTs are unknown, the method approximates it using the dataset $\mathcal{D}$. The resulting CBTs represent the characteristics of real and unknown CBTs. 
\begin{table}
	\caption{Evaluation of FN rate when new evidence arrives. Only weather, occlusion, road and reflection nodes are considered evidence nodes. Instead of grid map, overall prediction rate is calculated.}
	\begin{center}
		\begin{tabular}{|c|c|}
			\hline
			\textbf{Evidence}&\textbf{FN (Yes) accuracy} \\
			\hline
			Weather    & 75.4102\%	 \\
			\hline
			Occlusion  & 71.5815\% \\
			\hline
			Road & 76.1166\% \\
			\hline
			Reflection & 76.3445\% \\
			\hline
		\end{tabular}
		\label{tab:evaluation}
	\end{center}
\end{table}
Special care must be taken for tasks that are safety critical in nature. In the following, we discuss some of the assumptions that the parameter learning of BNs are based on which may challenge the robustness of the results.
\subsection{Representation of the Open Context}
The first and foremost assumption taken in any model is that it is considered as a good approximation of open context. In the specific case of BN, the structure represents the causal model and the CBT represents the relative occurrence as the approximation of open world phenomena for both data oriented or expert elicited CBTs. It may happen that not all the influencing factors are encoded in the BN structure and data does not represent the true relative frequency of phenomena. Dataset that does not well represent the open context may result in error prone PLMs and CPLMs.
\subsection{Rare Event Problem}
This concerns the well-known rare event occurrence frequency problem and its representation. This problem arises when there are important states of nodes which occur with lower frequency e.g. $illumination:tunnel$ $light$ is expected to occur with lower frequency than $illumination:day$. From the SOTIF standpoint, these states can also be safety critical. Evaluating robust CPLM for such states becomes challenging and is subject to perturbations. Such states can be artificially inserted in the data but the resultant marginalized probabilities will not be the true representation of the real world.\\
In this regard, a relative representation of each state frequency in the data explicitly modeled in the results can be a promising direction.   
\subsection{Training and Test Data}
Test dataset may inappropriately be segregated from the training data. Generally, test dataset should not be correlated with training dataset. However, in reality, highly correlated dataset is used because data is recorded at the same locations and it is recorded sequentially. This may lead to overestimated accuracy of the PLM.
\subsection{Data Abstraction and ODD Taxonomy}
Every scene is defined based on some abstraction. This is analogous to the data discretization problem in BN~\cite{vinh2012data}. Different abstractions may result in different maps e.g. a lower and more specific abstraction of $illumination$ node will be the values of light intensities instead of states such as $day$, a further lower abstraction might be taking a continuous light intensities distribution. Such distribution may result in different maps hence challenging the robustness of the results. Since these abstractions can be governed by operational design domain (ODD) taxonomies, a well-established ODD taxonomy can be used as the benchmark for data abstraction for analyses. Moreover, dynamic discretization can also be used in this regard~\cite{fenton2018risk}.
\section{Related Work} \label{ch:RelatedWork}
In recent years, extensive research has been done on the topic of SOTIF and scenario based safety of HAD vehicles~\cite{riedmaier2020survey}. However, to the best of authors' knowledge, existing approaches lack in the systematic identification, modeling, quantification and analysis of SOTIF relevant scenario factors. 
Berk et al.~\cite{berk2020} formalize the reliability-based validation of the environment perception for safe automated driving and discuss the associated challenges. The work focuses on the perception failure rate $\lambda_{per}$ and discusses the false negative (FN) and false positive (FP) as uncertainties. The implementation also provides qualitative and semi-quantitative analyses of sensor perception reliability. 
Ali et al.~\cite{ali2020} analyze the hazards arising due to variabilities in collaborative cyber physical systems (CPSs). Environmental, infrastructural, spatial and temporal variabilities are considered as factors causing uncertainties. They also develop a fault traceability graph to trace the faults considered by multiple hazard analyses in the collaborative CPSs with variability. 
Edward Schwalb~\cite{Schwalb2019AnalysisOS} provides a probabilistic framework for incrementally bounding the residual risk associated with autonomous drivers and enabling the quantifying progress. The work introduces continuous monitoring by autonomous driver for imminent hazards and selects actions that maximizes the time to materialization (TTM) of these hazards. The approach also enables implementing the continuous expansion of SOTIF through measurement of improvements from regressions using posterior probabilities.
Finally, Kramer et al.~\cite{Kramer2020} provide integrated method for safety assessment of automated driving functions, which covers the aspects of functional safety and SOTIF, including identification and quantification of hazardous scenarios. They also provide a functional insufficiency and causal chain analysis technique to identify and model SOTIF related hazards. Similar methodology is also presented in another literature~\cite{neurohr2021criticality}. However, the work provides a more theoretical view of the problem.  
\section{Conclusion and Future Work}
\label{ch:conclusion}
We presented a method to develop performance limitation maps (PLMs) as well as conditional performance limitations maps (CPLMs) under the scene model to study safety of the intended functionality (SOTIF). We identify the relevant triggering conditions, which are provided by experts and reasoned through data. The methodology encodes the parameter learning for Bayesian network (BN) for the implementation. 

This methodology particularly argues SOTIF under manageable effort. In its core, the provided methodology enables the analyst to identify performance limitations under various triggering conditions, their causal relations and limitations, conditioned on various phenomena critical under SOTIF. This further assists the analyst to establish mitigation strategies for the identified performance limitation under triggering conditions. In order to argue the adequacy of the approach, LIDAR performance was studied given a scene. The scene was modeled using a BN structure and parameter learning was performed using real world data to elicit conditional belief tables (CBTs).

We also evaluated the accuracy of learned BNs to demonstrate the predictive capabilities. We achieved roughly \textbf{75\%} when predicting the FN rate on the training data. We then discussed the robustness concerns of the safety methods in particular when data is used for parameter learning.

In future, we intend to explore how the robustness concerns of BNs can be addressed and mitigated. We particularly intend to provide methods focused on uncertainty measures and confidence intervals for CBTs and probabilities. Moreover, we also intend to model combined CPLMs for heterogeneous perception systems in order to identify and analyze common triggering conditions. The implementation can also be extended to test perception system based on same sensors with different governing algorithms. 

\bibliographystyle{IEEEtran}
\bibliography{IEEEabrv,references}
\end{document}